\documentclass[a4paper,12pt]{article}
\usepackage{latexsym,amssymb}
\usepackage{graphicx}
\usepackage{cite}
\usepackage[arrow,matrix,curve]{xy}
\usepackage{amsmath}

\global\arraycolsep=2pt 
\input{epsf} 

\def\s[#1,#2]{[#1\stackrel{{\displaystyle\star}}{,}#2]}

 1

\usepackage{amsthm}

\newcommand{\eq}{\begin{equation}}
\newcommand{\eqa}{\begin{eqnarray}}
\newcommand{\en}{\end{equation}}
\newcommand{\ena}{\end{eqnarray}}
\newcommand{\enn}{\nonumber \end{equation}}

\def\sk{\vskip .4cm}
\def\noi{\noindent}

\def\st {\star}
\def\f{{\rm{f}\,}}
\def\of{{\overline{{\rm{f}}\,}}}

\def\D/h{\widehat{\fmslash D}}

\def\om{\omega}
\def\Om{\Omega}
\def\al{\alpha}

\def\be{\beta}

\def\Ga{\Gamma}

\def\5bar{{\overline 5}}

\def\RR{{\mathcal R}}

\def\R{{R}}
\def\oR{{\overline{\R}}}

\def\ots{\otimes_\st}
\def\Oms{\Omega_\st}
\def\FF{\mathcal F}

\def\Xis{{\Xi_\st }}

\def\ll{{\mathcal L}}
\def\D{\Delta}
\def\TT{{\mathcal T}}

\def\nn{\nonumber}

\def\dd{{\triangledown}}
\def\dds{\triangledown^\st}
\def\rr{\mathsf{R}}
\def\ric{\mathsf{Ric}}
\def\tr{\mathsf{T}}

\def\ots{\otimes_\st}

\def\x{\mathsf{x}}
\def\y{\mathsf{y}}
\def\cc{\mathbb{C}}  
\def\r4{\mathbb{R}^4}  

\numberwithin{equation}{section}

\begin{document}

\begin{titlepage}
\rightline{DISTA-UPO/07}
\sk\sk
\begin{center}
{\bf{\LARGE{Noncommutative Gravity and the\\[.2em] $\st$-Lie algebra of diffeomorphisms
$\!^{^{_{_*}}}$}}}\\[.5em] 

\vskip 2.5em

{{\bf Paolo Aschieri}}

\vskip 1.5em
Centro Studi e Ricerche Enrico Fermi\\ Compendio Viminale, I-00184, Roma, 
Italy\\[.6em]
Dipartimento di Scienze e Tecnologie Avanzate\\
Universit\' a del Piemonte Orientale, and INFN - Sezione di Torino\\
Via Bellini 25/G 15100 Alessandria, Italy.\\
{E-mail: aschieri@to.infn.it}\\[1em]
\end{center}

\sk
\sk
\sk
\centerline{\bf Abstract}
\sk
\normalsize{
We construct functions and tensors
on noncommutative spacetime by systematically twisting the 
corresponding commutative structures. The study of the deformed 
diffeomorphisms (and Poincar\'e) Lie algebra allows 
to construct a noncomutative theory of gravity.
}\sk\sk
\sk\sk

\sk
\noi

{\footnotesize{\noi PACS: 02.40.Gh, 02.20.Uw, 04.20.-q,  11.10.Nx, 04.60.-m.~~~ 
}}

\sk\sk

\noi{\sl  This article is based on common work with Christian Blohmann, Marija 
Dimitrijevi\'c, Frank Meyer, Peter Schupp,  Julius Wess  \cite{G1, GR2}
and on \cite{GR3}.}
\sk
\noi-----------------------------------------------------------------\\
\noi {\footnotesize
\noi{$^*\,$Presented at the Erice International School of Subnuclear Physics 44th Course: homage to R. H. Dalitz ``The Logic of Nature, Complexity and New Physics: from quark-gluon plasma to superstrings, quantum gravity and beyond,''
Erice, Sicily, 29 August - 7 September 2006, and at the 
Second workshop and midterm meeting of the Marie Curie Research Training 
Network ``Constituents, Fundamental Forces and Symmetries of the Universe" 
Napoli, October 9-13, 2006}}
\sk


\sk\sk


\end{titlepage}\vskip.2cm

\newpage

\section{Introduction}

We present a differential geometry theory on noncommutative spacetime.
The simplest and most discussed example is canonical
noncommutative spacetime. 
There we have $x^\mu\st x^\nu-x^\nu\st x^\mu=i\theta^{\mu\nu}$. 
After introducing noncommutative functions and tensors we study their 
infinitesimal transformations laws and the corresponding deformed 
Lie algebra of infinitesimal diffeomorphisms (and as a particular case 
Poincar\'e transformations). 
Starting from these basic notions we can then define covariant derivatives
(so that we can  implement the principle of general covariance on 
noncommutative spacetime), and torsion and curvature tensors.
With these geometric tools we formulate noncommutative Einstein gravity. 
We here for simplicity consider noncommutative spacetime 
to be $\r4$ with canonical commutation relations \cite{G1,GR3}. Gravity on 
more general noncommutative manifolds, and its coordinate independent 
formulation is presented in \cite{GR2}.  
\sk
Among the motivations for the study of noncommutative field theories 
and gravity I would like to recall that in the passage from classical 
mechanics to quantum mechanics classical observables become noncommutative. 
Similarly we expect that in the passage from classical gravity to quantum 
gravity, gravity observables, i.e. spacetime itself, with its coordinates 
and metric structure, will become noncommutative. Thus by formulating Einstein 
gravity on noncommutative spacetime we may learn about quantum gravity.

Planck scale noncommutativity is further supported by Gedanken experiments 
that aim at probing spacetime structure at very small distances. 
They show that due to gravitational backreaction one cannot test 
spacetime at those distances. For example, 
in relativistic quantum mechanics the position of a particle can
be detected with a precision at most of the order of its 
Compton wave length $\lambda_C=\hbar/mc$.
Probing spacetime at infinitesimal distances implies an extremely 
heavy particle that in turn curves spacetime itself. 
When $\lambda_C$ is of the order of the Planck length, the spacetime
curvature radius due to the particle has the same order of magnitude and the 
attempt to measure spacetime structure beyond Planck scale fails.

This Gedanken experiment supports finite 
reductionism. 
It  shows that the description of spacetime
as a continuum of points (a smooth manifold) is an 
assumption no more justified at Planck scale. 
It is then natural to 
relax this assumption and conceive a noncommutative spacetime, 
where uncertainty relations and discretization naturally arise. 
In this way the dynamical feature of spacetime that prevents from testing 
sub-Plankian scales is explained by incorporating it at a deeper kinematical level.
A similar mechanism happens for example in the passage from 
Galilean to special relativity. Contraction of distances and 
time dilatation can be explained in Galilean relativity: 
they are a consequence of the interaction between ether and the 
body in motion. In special relativity they have become a kinematical feature.

\section{$\st$-Products and Twists}
The star product that implements the 
$x^\mu\st x^\nu-x^\nu\st x^\mu=i\theta^{\mu\nu}$ 
noncommutativity is given by  
\eq
(h\st g)(x)=e^{{i\over 2}\theta^{\mu\nu}{\partial\over \partial x^\mu}
{\partial\over \partial y^\nu}}h(x)g(y)|_{x=y}
\en
where $h$ and $g$ are arbitrary functions. 
This star-product between functions can be obtained from the usual 
pointwise product $(hg)(x)=h(x)g(x)$ via the action of a twist 
operator $\FF$ \cite{Oeckl, Chaichian}
\eq\label{starprodf}
f\st g:=\mu\circ \FF^{-1}(f\otimes g)~,
\en
where $\mu$ is the usual pointwise product between functions, 
$\mu(f\otimes g)=fg$, and the twist operator and its inverse are
\eq\label{MWTW}
\FF=e^{-{i\over 2}\theta^{\mu\nu}{\partial\over \partial x^\mu}
\otimes{\partial\over \partial x^\nu}}~,~~~
\FF^{-1}=e^{{i\over 2}\theta^{\mu\nu}{\partial\over \partial x^\mu}
\otimes{\partial\over \partial x^\nu}}~.
\en
Despite the indices $^{\mu~\nu}$ notation, we will consistently consider the entries
$\theta^{\mu\nu}$ of the antisymmetric matrix $\theta$ as fundamenal dimensionful constants, like $c$ or $\hbar$.
In particular the deformed spacetime symmetries we consider will leave 
invariant the $\theta$ matrix.  
The point is that the exponent of $\FF$,  
$$\theta^{\mu\nu}{\partial\over \partial x^\mu}
\otimes{\partial\over \partial x^\nu}$$ is not 
the Poisson tensor associated to the $\st$-product.
The difference lies in the tensorproduct $\otimes$. 
The Poisson tensor is
\eq\label{Poisson}
\theta^{\mu\nu}
{\partial\over \partial x^\mu}
\otimes_A{\partial\over \partial x^\nu}
\en
where we have explicitly written that the tensorproduct is over the algebra $A=Fun(\r4)$ of functions on spacetime.
On the other hand the tensorproduct in $\FF$ is over the complex numbers, 
we should write
$$ 
\FF=e^{-{i\over 2}\theta^{\mu\nu}{\partial\over \partial x^\mu}\:
\otimes{_\cc}\:{\partial\over \partial x^\nu}}~.
$$ 
That is why $\theta^{\mu\nu}$ in $\FF$ is not a tensor but a set 
of constants. In this respect, a better notation for $\FF$ is
\eq
\FF=e^{{-i\over 2}\theta^{a b}{X_a}
\otimes X_b}~.
\en
where $a,b=1,...4$ and $X_1={\partial\over \partial x^1}\,,\ldots\,
X_4={\partial\over \partial x^4}$, are { globally} 
defined vectorfields on spacetime. 
\sk
We shall frequently write (sum over $\al$ understood)
\eq\label{Fff}
\FF=\f^\al\otimes\f_\al~~~,~~~~\FF^{-1}=\of^\al\otimes\of_\al~,
\en
so that
\eq\label{fhfg}
f\st g:=\of^\al(f)\of_\al(g)~.
\en
Explicitly we have
\eq
\FF^{-1}=e^{{i\over 2}\theta^{\mu\nu}{\partial\over \partial x^\mu}
\otimes{\partial\over \partial x^\nu}}
=\sum {1\over{n!}}\left( i\over 2\right)^n\theta^{\mu_1\nu_1}\ldots\theta^{\mu_n\nu_n}
\partial_{\mu_1}\ldots\partial_{\mu_n}\otimes
{}\partial_{\nu_1}\ldots\partial_{\nu_n}=\of^\al\otimes\of_\al~,\label{faexp}
\en
so that $\al$ is a multi-index.
We also introduce the universal $R$-matrix
\eq
\RR:=\FF_{21}\FF^{-1}~\label{defUR}
\en
where by definition $\FF_{21}=\f_\al\otimes \f^\al$.
In the sequel we use the notation 
\eq
\RR=\R^\al\otimes\R_\al~~~,~~~~~~\RR^{-1}=\oR^\al\otimes\oR_\al~.
\en
In the present case we simply have $\RR=\FF^{-2}$ but for
more general twists this is no more the case.
The $R$-matrix measures the noncommutativity of the $\star$-product.
Indeed it is easy to see that 
\eq\label{Rpermutation}
h\st g=\oR^\al(g)\st\oR_\al(h)~.
\en
The permutation group in noncommutative space is naturally represented by 
$\RR$. Formula (\ref{Rpermutation}) says that the $\st$-product is 
$\RR$-commutative in the sense that if we permute (exchange) two functions 
using the $R$-matrix action then the result does not change.
\sk
\noi{\sl Note: } The class of $\st$-products that can be obtained from a twist 
$\FF$ is quite rich, (for example we can obtain star products that give  the commutation relations $x\st y=q y \st x$ in two or more dimensions). Moreover we can consider twists and $\st$-products on 
arbitrary manifolds not just on $\r4$. For example, given a set of 
mutually commuting 
vectorfields $\{X_a\}$  ($a=1,2,\ldots n$) on a $d$-dimensional manifold $M$, 
we can consider the twist 
\eq\label{betternotation}
\FF=e^{{-i\over 2}\theta^{a b}{X_a}
\otimes X_b}~.
\en
Another example is 
$\FF=e^{{1\over 2}H\otimes ln(1+\lambda E)}$ where the vectorfields 
$H$ and $E$ satisfy $[H,E]=2E$. 
In these cases too the $\st$-product defined via (\ref{starprodf})
is associative and properly normalized in the sense that for any 
function 
$h$ we have $h\st 1=1\st h=h$. In general an element $\FF$ is a 
twist if it is invertible, if it satisfies a cocycle condition 
and if it is properly normalized \cite{Drinfeld1}.  
The cocycle and the normalization conditions imply 
associativity of the $\st$-product and the normalization
$h\st 1=1\st h=h$. A deformed gravity theory based on this class of 
$\st$-products can also be constructed \cite{GR2}.

\section{Vectorfields and Tensorfields}
We now use the twist to  
deform the commutative geometry on spacetime 
into the twisted noncommutative one. 
The guiding principle is the one used to deform the product of 
functions into the $\st$-product of functions. Every time we have 
a bilinear map $$\mu\,: X\times Y\rightarrow Z~~~~~~~~~~~~~~~~~~$$
where $X,Y,Z$ are vectorspaces, 
and where there is an action of $\FF^{-1}$ on $X$ and $Y$
we can combine this map with the action of the twist. In this way
we obtain a deformed version $\mu_\st$ of the initial bilinear map $\mu$:
\eqa
\mu_\st:=\mu\circ \FF^{-1}~,\label{generalpres}&~~~~~~~~~~~~~~&
\ena
{\vskip -.8cm}
\eqa
{}~~~~~~~~~~~~~\mu_\st\,:X\times  Y&\rightarrow& Z\nn\\
(\x, \y)\,\, &\mapsto& \mu_\st(\x,\y)=\mu(\of^\al(\x),\of_\al(\y))\nn~.
\ena
The $\st$-product on the space of functions is recovered setting 
$X=Y=A=Fun(M)$. 
We now study the case of vectorfields and tensorfields.
\sk
\noi {\it Vectorfields $\Xi_\st$}. We deform the product 
$\mu : A\otimes \Xi\rightarrow \Xi$ between 
the space $A=Fun(M)$ of functions on spacetime $M$ and vectorfields. 
A generic vectorfield is 
$v=v^\nu\partial_\nu$. Partial derivatives acts on vectorfields via the 
Lie derivative action
\eq\label{onlyconst}
\partial_\mu(v)=[\partial_\mu,v]=\partial_\mu(v^\nu)\partial_\nu~.
\en
According to (\ref{generalpres}) the product 
$\mu : A\otimes \Xi\rightarrow \Xi$
is deformed into the product 
\eq
h\st v=\of^\al(h) \of_\al(v)~.
\en
Since $\FF^{-1}=e^{{i\over 2}\theta^{\mu\nu}\partial_\mu
\otimes \partial_\nu}$,
iterated use of (\ref{onlyconst})  gives
\eq
h\st v=\of^\al(h) \of_\al(v)
=\of^\al(h)\of_\al(v^\nu) \partial_\nu=
(h\st v^\nu)\partial_\nu ~.
\en
It is then easy to see that  $h\st (g\st v)=(h\st g)\st v$, i.e. that
the $\st$-multiplication between functions and vectorfields is consistent with 
the $\st$-product of functions. We denote the space of vectorfields with this 
$\st$-multiplication  by $\Xi_\st$. 
As vectorspaces $\Xi=\Xi_\st$, but $\Xi$ is an $A$-module while $\Xi_\st$ is 
an $A_\st$-module.
\sk
\noi {\it Tensorfields {$\TT_\st$}}.  Tensorfields form an algebra with 
the tensorproduct $\otimes$ (over the algebra of functions). We define $\TT_\st$ to be the noncommutative 
algebra of tensorfields. As vectorspaces $\TT=\TT_\st$; the noncommutative and associative
tensorproduct is obtained by applying (\ref{generalpres}):
\eq\label{defofthetensprodst}
\tau\otimes_\st\tau':=\of^\al(\tau)\otimes \of_\al(\tau')~.
\en
There is a natural action of the permutation group on undeformed tensorfields:
$$\tau\otimes\tau'\stackrel{\sigma}{\longrightarrow} \tau'\otimes\tau~.$$ 
In the deformed case it is the $R$-matrix that provides a representation 
of the permutation group on $\st$-tensorfields:
$$\tau\otimes_\st\tau'\stackrel{\sigma_{_\RR}}{\longrightarrow}
\oR^\al(\tau')\otimes_\st \oR_\al(\tau)~.$$ 
It is easy to check that, consistently with $\sigma_\RR$ 
being a representation of the permutation group, we have $(\sigma_\RR)^2=id$.

\sk
If we consider the local coordinate expression of two tensorfields,
for example of the type
$$
\tau =\tau^{\mu_1,...\mu_m}\partial_{\mu_1}\otimes_\st\ldots\partial_{\mu_m}\\
$$
$$\tau' =\tau'^{\nu_1,...
\nu_n}\partial_{\nu_1}\otimes_\st\ldots\partial_{\nu_n}
$$
then their $\st$-tensor product is  
\eq
\tau\otimes_\st\tau'=
\tau^{\mu_1,...\mu_m}\st\tau'^{\nu_1,...\nu_n}
\partial_{\mu_1}\otimes_\st\ldots\partial_{\mu_m}
\otimes_\st\partial_{\nu_1}\otimes_\st\ldots\partial_{\nu_n}~.
\en
\section{$\st$-Lie algebra of Diffeomorphisms}
In the commutative case we have two derivations that map tensors to tensors 
(of the same type):
the Lie derivative $\ll_v$ and the covariant derivative $\dd_v$ 
along the vectorfield $v$. If we act on functions 
(tensors of type $0$) Lie derivative and covariant derivative coincide 
and are the usual action of the vectorfield $v$ on a function $h$. 
The result of this action is 
the variation of the function $h$ under the infinitesimal transformation
generated by $v$, $$\delta_vh=\ll_vh=v(h)
~.$$
In this section we study the $\st$-action of a vectorfield on a function, 
i.e. we deform the notion of Lie derivative (infinitesimal 
diffeomorphism), and study the deformed Lie algebra of these infinitesimal 
diffeomorphisms. Once this deformed derivation is understood then the notion 
of deformed covariant derivative will naturally emerge, and the construction 
of the Rimemann curvature tensor, the Ricci tensor and
the curvature scalar will follow.  These tensors will transform
covariantly under infinitesimal diffeomorphisms.
\sk
The $\st$-Lie derivative on the space of functions $A_\st$ 
is obtained following the general 
rule (\ref{generalpres}).
We combine the usual Lie derivative on functions 
$\ll_uh=u(h)$ with the twist $\FF$ 
\eq\label{stliederact}
\ll^\st_u(h):=\of^\al(u)(\of_\al(h))~.
\en
By recalling that every vectorfield can be written as 
$u=u^\mu\st\partial_\mu=u^\mu\partial_\mu$ we have
\eqa\nn
\ll^\st_u(h)&=&\of^\al(u^\mu\partial_\mu)(\of_\al(h))=\of^\al(u^\mu)\,\partial_\mu(\of_\al(h))\\[.3em]
&=&
u^\mu\st\partial_\mu(h)~,\label{exlu}
\ena
where in the second equality we have considered the explicit expression  (\ref{faexp}) of $\of^\al$ in terms of partial derivatives, and we have iteratively used the property 
$[\partial_\nu,u^\mu\partial_\mu]=\partial_\nu(u)\,\partial_\mu$. In the last equality we have used that the partial derivatives contained in $\of_\al$ commute with the partial derivative $\partial_\mu$.

The differential operator $\ll^\st_u$ satisfies the deformed  Leibniz rule 
\eq\label{defL}
\ll^\st_u(h\st g)=\ll_u^\st(h)\st g + 
\oR^\al(h)\st \ll_{\oR_\al(u)}^\st(g)~.
\en
This deformed Leibnitz rule is intuitive: in the second addend we have 
exchanged the order of $u$ and $h$, and this is achieved by 
the action of the $R$-matrix that as observed in Section 2 and 3 
provides a representation of the permutation group. 
A proof of (\ref{defL}) is not difficult
\eqa
\ll^\st_u(h\st g)=
u^\mu\st \partial_\mu(h\st g)&=&u^\mu\st \partial_\mu(h)\st g 
+ u^\mu\st h\st\partial_\mu(g)\nn\\[.3em]&=&
\ll^\st_u(h)\st g 
+ \oR^\al(h)\st \oR_\al(u^\mu)\st\partial_\mu(g)\nn\\[.3em]
&=&\ll_u^\st(h)\st g 
+ \oR^\al(h)\st \ll_{\oR_\al(u)}^\st(g)~.
\ena

From (\ref{exlu}) it is also immediate to check the compatibility condition
\eq
\ll_{f\st u}^\st(h)=f\st\ll_u^\st(h)~,
\en
that shows that the action $\ll^\st$ on functions is the one compatible with the
$A_\st$ module structure of vectorfields.
\sk

In the commutative case the commutator of two vectorfields 
is again a 
vectorfield, we have the Lie algebra of vectorfields. In this 
$\st$-deformed case we have a similar situation.
We first calculate 
\eq
\ll^\st_u\ll^\st_v(h)=\ll^\st_u(\ll^\st_v(h))
=u^\mu\st\partial_\mu(v^\nu)\st \partial_\nu (h)+
u^\mu\st v^\nu\st\partial_\nu\partial_\mu(h)
\nn\en
Then instead of considering the composition $\ll^\st_v\ll^\st_u$ 
we consider 
$\ll^\st_{\oR^\al(v)}\ll^\st_{\oR_\al(u)}$, indeed the usual commutator is 
constructed permuting (transposing) the two vectorfields, 
and we have 
just remarked that the action of the permutation group in the 
noncommutative case is obtained using the $R$-matrix.
We have 
\eqa
\ll^\st_{{\oR^\al}(v)}\ll^\st_{\oR_\al(u)}(h)&=& 
\ll^\st_{{\oR^\al}(v)}(\oR_\al(u^\mu)\st\partial_\mu h)= 
{\oR^\al}(v^\nu)\st\partial_\nu(\oR_\al(u^\mu)\st\partial_\mu h)\nonumber\\
&=&{\oR^\al}(v^\nu)\st \oR_\al(\partial_\nu u^\mu)\st\partial_\mu h
+{\oR^\al}(v^\nu)\st\oR_\al(u^\mu)\st\partial_\nu \partial_\mu h
\nonumber
\ena
In conclusion
\eq
\ll^\st_u\ll^\st_v- \ll^\st_{\oR^\al}(v)\ll^\st_{\oR_\al(u)}=\ll_{[u,v]_\st}
\en
where we have defined the new vectorfield
\eq\label{stbrack}
[u,v]_\st := (u^\mu\st\partial_\mu v^\nu)
\partial_\nu-(\partial_\nu u^\mu\st v^\nu)\partial_\mu~.
\en
The bracket $[~~,~~]_\st$ is a bilinear map
\eqa
[~~,~~]_\st~:~\Xi_\st \times\Xi_\st &\rightarrow& \Xi_\st\nonumber\\
                (u,v)&\mapsto & [u,v]_\st
\ena
and the space of vectorfields equipped with this bracket becomes 
the $\st$-Lie algebra of vectorfields $\Xi_\st$. 
It is easy to see that the bracket  $[~~,~~]_\st $ 
has the $\st$-antisymmetry property
\eq\label{sigmaantysymme}
[u,v]_\st =-[\oR^\al(v), \oR_\al(u)]_\st~ .
\en
$\st$-Jacoby identities can be proven as well
\eq\label{stJac}
[u,[v,z]_\st ]_\st =[[u,v]_\st ,z]_\st  
+ [\oR^\al(v), [\oR_\al(u),z]_\st ]_\st ~.
\en
\sk
In conclusion we have constructed the deformed Lie algebra of vectorfields 
$\Xi_\st$. As vectorspaces $\Xi=\Xi_\st$, but $\Xi_\st$ is a $\st$-Lie algebra.
We previously constructed $\Xi_\st$ as an $A_\st$-module. 
The compatibility between these two structures reads
$$[u,h\st v]_\st=\ll^\st_u(h)\st v + \oR^\al(h)\st [{\oR_\al(u)},v]_\st~.$$

We stress that a $\st$-Lie algebra is not a generic name for a 
deformation of a Lie algebra. Rather it is a quantum Lie algebra in the
sense of \cite{Woronowicz}, see also \cite{AC, SWZ},\cite{GR3} and
\cite{Modave}. 
In this respect the deformed Leibnitz rule (\ref{defL}), that states that 
only vectorfields (or the identity) can act on the second argument 
$g$
in $h\st g$
(no higher order differential operators are allowed on $g$) is of fundamental 
importance, and later on it will be a key ingredient for the definition of a 
covariant derivative along a generic vectorfield. 
Another main property
of quantum Lie algebras, and  of the quantum Lie algebra of 
infinitesimal diffeomorphisms we have presented, is that  it can be shown 
that the bracket $[u~v]_\st$ is the $\st$-adjoint action of $u$ on $v$. 
Had we chosen an exactly antisymmetric bracket of vectorfields, 
we would have lost this geometric property. 

\sk
A general comment on the approach adopted
is now in order.
Given the deformation $Fun_\st(M)$ of the algebra of functions $Fun(M)$, 
we can
\begin{itemize}
\item consider the derivations of $Fun_\st(M)$, i.e.
the infinitesimal transformations of $Fun_\st(M)$ that satisfy the usual
Leibnitz rule, 
$v(h\st g)=v(h)\st g+ h\st v(g)$. 
As is easily seen expanding in power series of $\theta^{\mu\nu}$,
these maps are only the vectorfields that leave invariant the 
Poisson tensor (\ref{Poisson}). Thus while in the
commutative case any vectorfield 
is a derivation, in the deformed case the 
space of derivations is smaller. This viewpoint for our pourposes 
is too 
restrictive, for example infinitesimal Poincar\'e transformations are not 
derivations. In this approach we have that Poincar\'e invariance is 
spontaneously broken by the presence of $\theta^{\mu\nu}$. 
\sk
\item consistently deform the notion of derivation so that
to any infinitesimal 
transformation of $Fun(M)$ there correspond one and only one deformed 
infinitesimal derivation. This is what we have achieved with the map 
$\ll_v\rightarrow \ll^\st_v$, where $\ll_v^\st$ satisfies the deformed 
Leibnitz rule (\ref{defL}). This is the quantum groups and quantum spaces 
approach \cite{WessZumino, Woronowicz, AC, SWZ}.
The bonus of this approach is that instead of dealing 
with a spontaneously broken diffeomorphisms (or Poincar\'e) symmetry
we have an unbroken quantum  diffeomorphisms (or Poincar\'e) symmetry. 
In this way we retain a symmetry property that is as strong as the 
one of commutative spacetime.
\end{itemize}
\sk
\noi
{\bf Deformed Poincar\'e Lie algebra}\\
The quantum or twisted approach in the case of Poincar\'e symmetry 
of canonical noncommutative spacetime 
was considered in 
\cite{Wess, Chaichian, Oeckl}, where the twisted Poincar\'e 
Hopf algebra is presented (see also  
\cite{Koch, Gonera:2005hg}). 
The description of this twised symmetry in terms  
of $\st$-Poincar\'e Lie algebra, i.e.  
the quantum Lie algebra of the Poincar\'e Hopf algebra, 
is in \cite{GR3}.
Using the bracket (\ref{stbrack}), we have that the 
infinitesimal generators 
\eq
\label{representation}  {P}_\mu =i\partial_\mu \ ,\ {M}_{\mu\nu}
=i(x_\mu\partial_\nu - x_\nu\partial_\mu)\ ,
\en
close the
$\st$-Poincar\'e
Lie algebra 
\eqa\label{LiePoincst} [P_\mu,P_\nu]_{\st}&=&0\ ,\nn\\[.3em]
[P_\rho , M_{\mu\nu}]_{\st}&=&i(\eta_{\rho\mu}P_\nu-\eta_{\rho\nu}P_\mu)
\,,\nn\\[.3em]
[M_{\mu\nu},M_{\rho\sigma}]_{\st}&=&
-i(\eta_{\mu\rho}M_{\nu\sigma}-\eta_{\mu\sigma}M_{\nu\rho}
-\eta_{\nu\rho}M_{\mu\sigma}+\eta_{\nu\sigma}M_{\mu\rho})\ .
\ena
In this particularly simple case of canonical noncommutativity 
it happens that
the structure constants are the same as in the undeformed case,
however the $\st$-bracket differs from the commutator,  
$\ll^\st_{[M_{\mu\nu},M_{\rho\sigma}]_{\st}}\not=
\ll_{M_{\mu\nu}}^\st\ll^\st_{M_{\rho\sigma}}-\ll^\st_{M_{\rho\sigma}}
\ll^\st_{ M_{\mu\nu}}$.
The deformed Leibnitz rule of these derivations, according to (\ref{defL}), 
reads
$$
\ll^\st_{M_{\mu\nu}}(h\st g)=\ll^\st_{M_{\mu\nu}}(h)\st g+
h\st\ll^\st_{M_{\mu\nu}}(g)
-i\theta^{\al\be}\eta_{\beta\mu}\partial_\al(h)\st\partial_\nu(g)
+i\theta^{\al\be}\eta_{\beta\nu}\partial_\al(h)\st\partial_\mu(g)~.
$$

From  (\ref{stJac}) or immediately from (\ref{LiePoincst}) we obtain 
the $\st$-Jacoby identities:
\eq
[t \,,[t',t'']_{_\st} ]_{_\st} +[t' \,,[t'',t]_{_\st} ]_{_\st} 
+ [t'' \,,[t,t']_{_\st} ]_{_\st} =0~,
\en
for all elements $t,t',t''$ of the $\st$-Poincar\'e Lie algebra. 
\sk

\section{Covariant Derivative, Torsion and Curvature}\label{covderivative}
The noncommutative differential geometry set up 
in the previous section allows
to develop the formalism of covariant
derivative, torsion, curvature 
and Ricci tensors just by following the usual classical formalism.

We define a $\st$-covariant derivative 
$\dd^\star_u$ along the vector field $u\in \Xi$
to be a linear map $\dds_u:\Xis\rightarrow\Xis$ such that
for all $u,v,z\in\Xi_\st,~ h\in A_\st$:
\eqa
&&\dd_{u+v}^{\star}z=\dd_{u}^{\star}z+\dd_{v}^{\star}z~,\\[.35cm]
&&\dd_{h\star u}^{\star}v=h\star\dd_{u}^{\star}v~,\label{ddal}\\[.35cm]
&&\dd_{u}^{\star}(h\star v)
\,=\,\mathcal{L}_u^{\star}(h)\star v+
\oR^\al(h)\st\dd^\st_{\oR_\al(u)}v\label{ddsDuhv}
\ena
where in the last line we have used the same deformed Leibnitz rule 
that appears in (\ref{defL}). Epression (\ref{ddsDuhv}) is well defined because
$\oR_\al(u)$ 
is again a vectorfield.

The covariant derivative is extended to tensorfields using again the deformed 
Leibniz rule
$$\dds_u(v\otimes_\st z)= \dds_{u}(v)\ots z + \oR^\al(v)\ots 
\dds_{\oR_\al(u)}(z)\,\,.
$$

\sk
The (noncommutative) connection coefficients 
${\Gamma_{\mu\nu}}^\sigma$ are given by 
\eq
\dds_{\mu}\partial_\nu=
{\Gamma_{\mu\nu}}^\sigma\st\partial_\sigma=
{\Gamma_{\mu\nu}}^\sigma\,\partial_\sigma~,
\en
where  $\dds_\mu=\dds_{\partial_\mu}$.
Let 
$z=z^\mu\st\partial_\mu$,
$u=u^\nu \st\partial_\nu$, then (\ref{ddal}) and (\ref{ddsDuhv}) imply
\eq
\dds_z u=z^\mu\st\partial_\mu(u^\nu)\,\partial_\nu+
z^\mu\st u^\nu\st \Gamma_{\mu\nu}{}^\sigma\,\partial_\sigma\, .
\en
Similarly the covariant derivative on 1-forms is given by
\eq
\dds_{\mu} (\om_\rho  dx^\rho)
=\partial_\mu(\om_\rho)\,dx^\rho-
\Ga_{\mu\rho}{}^\nu\st \om_\nu\: dx^\rho ~
\en
and  $\dds_z=z^\mu\st\dds_\mu\,$.
\sk
The torsion $\tr$ and the curvature $\rr$ associated to
a connection $\dd^\st$ are the linear maps  
$\tr:\Xis \times \Xis\rightarrow\Xis$, and 
$\rr^\star : \Xis\times \Xis\times\Xis\rightarrow\Xis$ defined by
\eqa
\tr(u,v)&:=&\dd_{u}^{\star}v-\dd_{\oR^{\alpha}(v)}^{\star}\oR_{\alpha}(u)
-[u,v]_{\star}~,\\[.2cm]
\rr(u,v,z)&:=&\dd_{u}^{\star}\dd_{v}^{\star}z-
\dd_{\oR^\al{(v)}}^{\star}\dd_{\oR_\al(u)}^{\star}z-\dds_{[u,v]_\st} z~,
\ena
for all $u,v,z\in\Xis$.
{}From  the $\st$-antisymmetry property of the bracket $[~,~]_\st$, 
see (\ref{sigmaantysymme}), it
easily follows that 
the torsion $\tr$ and the curvature $\rr$ have the following 
$\st$-antisymmetry property
\eqa\label{Tantysymm}
\tr(u,v)&=&-\tr(\oR^\al(v),\oR_\al(u))~,\nn\\[.3em]
\rr(u,v,z)&=&-\rr(\oR^\al(v),\oR_\al(u),z)~.\nn
\ena
The presence of the $R$-matrix in the definition of torsion and curvature insures 
that  $\tr$ and $\rr$  are left $A_\st$-linear maps \cite{GR2}, i.e.
$$
\tr(f\star u,v)=f\star \tr(u,v)~
~~,~~~\tr(\partial_\mu,f\star v)= f\st\tr(\partial_\mu ,v)
$$
(for any index $\mu$), and similarly for the curvature.
The $A_\st$-linearity of $\tr$ and $\rr$ insures that we
have a well defined  torsion tensor and  curvature  tensor.

One can also prove (twisted) first and second Bianchi identities
\cite{GR2}.
\sk
The coefficients  $\tr_{\mu\nu}{}^\rho$  and $\rr_{\mu\nu\rho}{}^\sigma$
with respect to the partial derivatives basis $\{\partial_\mu\}$ are defined by
\eq
\tr(\partial_\mu,\partial_\nu)=\tr_{\mu\nu}{}^\rho\partial_\rho~~~,~~~~~~
\rr(\partial_\mu,\partial_\nu,\partial_\rho)=
\rr_{\mu\nu\rho}{}^\sigma\partial_\sigma
\en 
and they explicitly read
\eqa
\tr_{\mu\nu}{}^\rho&=&\Ga_{\mu\nu}{}^\rho-\Ga_{\nu\mu}{}^\rho~~,\nn\\[.3em]
\rr_{\mu\nu\rho}{}^\sigma&=&\partial_\mu\Ga_{\nu\rho}{}^\sigma-
\partial_\nu\Ga_{\mu\rho}{}^\sigma
+\Ga_{\nu\rho}{}^\beta\st\Ga_{\mu\beta}{}^\sigma-
\Ga_{\mu\rho}{}^\beta\st\Ga_{\nu\beta}{}^\sigma~.
\ena

As in the
commutative case the Ricci tensor is a contraction of the curvature
tensor, 
\eq
\ric_{\mu\nu}={\rr_{\rho\mu\nu}}^\rho.
\en
A definition of the Ricci tensor that is independent from the $
\{\partial_\mu\}$ basis is also possible.

\section{Metric and Einstein Equations}\label{riemgeo}
In order to define a $\st$-metric we need to define $\st$-symmetric 
elements in $\Oms\ots\Oms$ where $\Oms$ is the space of 1-forms. 
Recalling that
permutations are implemented with the $R$-matrix 
we see that  $\st$-symmetric elements are of the form
\eq
\omega\otimes_\st\omega'
+\oR^\al(\omega')\otimes_\st \oR_{\al}(\omega)~.
\en
In particular any symmetric tensor in
$\Om\otimes\Om\,$,
\eq
g=g_{\mu\nu}dx^\mu\otimes dx^\nu~,
\en
$g_{\mu\nu}=g_{\nu\mu}$, is also a $\st$-symmetric tensor in 
 $\Oms\ots\Oms$ because
\eq
g=g_{\mu\nu}dx^\mu\otimes dx^\nu=g_{\mu\nu}\st dx^\mu\otimes_\st dx^\nu~
\en
and the action of the $R$-matrix is the trivial one on $dx^\nu$. 
We denote by $g^{\st \mu\nu}$ the star inverse of $g_{\mu\nu}$,
\eq
g^{\st \mu\rho}\st g_{\rho\nu}
=g_{\nu\rho}\st g^{\st \rho\mu}=\delta_\nu^\mu~.
\en
The metric $g_{\mu\nu}$ can be expanded order by order in the
noncommutative parameter $\theta^{\rho\sigma}$.
Any commutative metric is also a noncommutative metric, indeed the 
$\st$-inverse metric can be constructed order by order in the noncommutativity
parameter. Contrary to \cite{MadoreFC}, 
we see that in our approach
there are infinitely many metrics compatible with 
a given noncommutative differential geometry, noncommutativity 
does not single out a preferred metric.

\sk
A connection that is metric compatible is a connection that for any vectorfield $u$ satisfies,
$\dds_u g=0$, this is equivalent to the equation
\eq
\dds_\mu g_{\rho\sigma}-\Ga_{\mu\rho}{}^\nu \st g_{\nu\sigma} -\Ga_{\mu\sigma}{}^\nu \st g_{\rho\nu}=0 ~.
\en
Proceeding as in the commutative case we obtain that there is a unique torsion 
free metric compatible connection \cite{G1}. It is given by
\eq
\Ga_{\mu\nu}{}^\rho={1\over 2}(\partial_\mu g_{\nu\sigma}+
\partial_\nu g_{\sigma\mu}-\partial_\sigma g_{\mu\nu})\st g^{\st \sigma\rho}
\en
\sk

We now construct the curvature tensor and the Ricci tensor using this 
uniquely defined connection. Finally the noncommutative Einstein equations 
(in vacuum) are
\eq\label{Einstein}
\ric_{\mu\nu}=0
\en
where the dynamical field is the metric $g$.
\sk
\sk\sk
\noi{\bf{\large Acknowledgements}}
\sk
\noi I acknowledge the stimulating and fruitful atmospheres 
of the Erice International school of Subnuclear Physics, 44t course (2006) 
and of the Marie Curie RTN workshop ``Constituents, fundamental forces and
    symmetries in the universe'', Napoli (2006). Partial support from 
Ettore Majorana Foundation and Centre for Scientific Studies, Erice, is
kindly acknowledged. Partial support from the European Community's Human 
Potential Program under contract MRTN-CT-2004-005104 and from the
Italian MIUR under contract PRIN-2003023852 is also acknowledged.

\end{document}